\documentclass[reprint,aps,prl,superscriptaddress,showpacs]{revtex4-1}
\usepackage{hyperref}
\usepackage{graphicx}
\usepackage{amssymb,amsmath}
\usepackage{color}

\begin{document}

\title{Scattering from Charged Black Holes and Supergravity}

\author{Lu\'is C. B. Crispino}
 \email{crispino@ufpa.br}
  \affiliation{Faculdade de F\'isica, Universidade Federal do Par\'a,
66075-110, Bel\'em, Par\'a, Brazil}

\author{Sam R. Dolan}
\email{s.dolan@sheffield.ac.uk}
\affiliation{Consortium for Fundamental Physics,
School of Mathematics and Statistics,
University of Sheffield, Hicks Building, Hounsfield Road, Sheffield S3 7RH, United Kingdom}

\author{Atsushi Higuchi}
 \email{atsushi.higuchi@york.ac.uk}
  \affiliation{Department of Mathematics, University of York,
Heslington,
York YO10 5DD, United Kingdom}

\author{Ednilton S. de Oliveira}
 \email{ednilton@pq.cnpq.br}
 \affiliation{Faculdade de F\'isica, Universidade Federal do Par\'a,
66075-110, Bel\'em, Par\'a, Brazil}


\begin{abstract}
The scattering cross sections of the unpolarized electromagnetic and gravitational waves
are shown to be equal for the extreme Reissner-Nordstr\"om black hole using $N=2$ supergravity.
The conversion cross sections between the unpolarized electromagnetic and gravitational waves are also shown
to coincide.  The gravitational and electromagnetic scattering cross sections are computed
numerically for Reissner-Nordstr\"om black holes for several charge-to-mass ratios and the coincidence between
the two scattering cross sections for the extremal case is confirmed.
\end{abstract}

\maketitle

\emph{Introduction:} At the center of our galaxy, an intense compact radio source, Sagittarius A*, is associated 
with a supermassive black hole of mass $M \sim 4.31 \times 10^6 M_\odot$~\cite{Gillessen:2008qv}. 
It is thought that most galaxies have a supermassive core, with a close link between the properties 
of the black hole and the galaxy itself. Each galaxy also plays host to myriad black holes formed by stellar collapse. 
Clearly, black holes are a fundamental constituent of the universe, whose exotic properties are still being understood.

According to the no-hair conjecture, 
when a black hole in electrovacuum is perturbed, 
it settles back into a Kerr-Newman configuration characterized by just three numbers:
mass $M$, angular momentum $J$ and electric charge $Q$~\cite{Chrusciel:2012jk}. 
Observations suggest astrophysical black holes have significant 
angular momentum~\cite{Risaliti:2013cga}; it is not known whether they are also endowed with charge
though a Schwinger process would rapidly 
decrease the charge unless $M \gtrsim 10^5 M_\odot$~\cite{Gibbons:1975kk}.
In the extremal limit $(J/M)^2 + Q^2 \rightarrow M^2$ (with natural units $G=c=1$), the surface gravity of the 
horizon, and thus the Hawking temperature, approaches zero. Extremal black holes may be regarded as solitons of 
Einstein-Maxwell theory~\cite{Gibbons:1982fy}. 

In this Letter we consider planar electromagnetic and gravitational waves interacting with Reissner-Nordstr\"om 
(RN) black holes, with $J=0$, $0 \le Q \le M$. 
A wave impinging upon a black hole will be partially absorbed and 
scattered, and the polarized flux will bear the imprint of the black hole. For example, strong-field scattering near the 
black hole light-ring generates a distinctive `spiral scattering' interference, with a `glory' on-
axis~\cite{Crispino_2009:prl102_231103}. When the black hole possesses charge, the electromagnetic 
and gravitational perturbations are mixed together: an incident electromagnetic wave will generate a gravitational 
wave, and vice versa~\cite{Olson_1974:prl33_1116}. Furthermore, the helicity of the incident wave is not conserved, 
leading to novel signatures~\cite{Crispino_2014:prd90_064027}. 

\begin{figure}[htb!]
 \centering
 \includegraphics[width=\columnwidth]{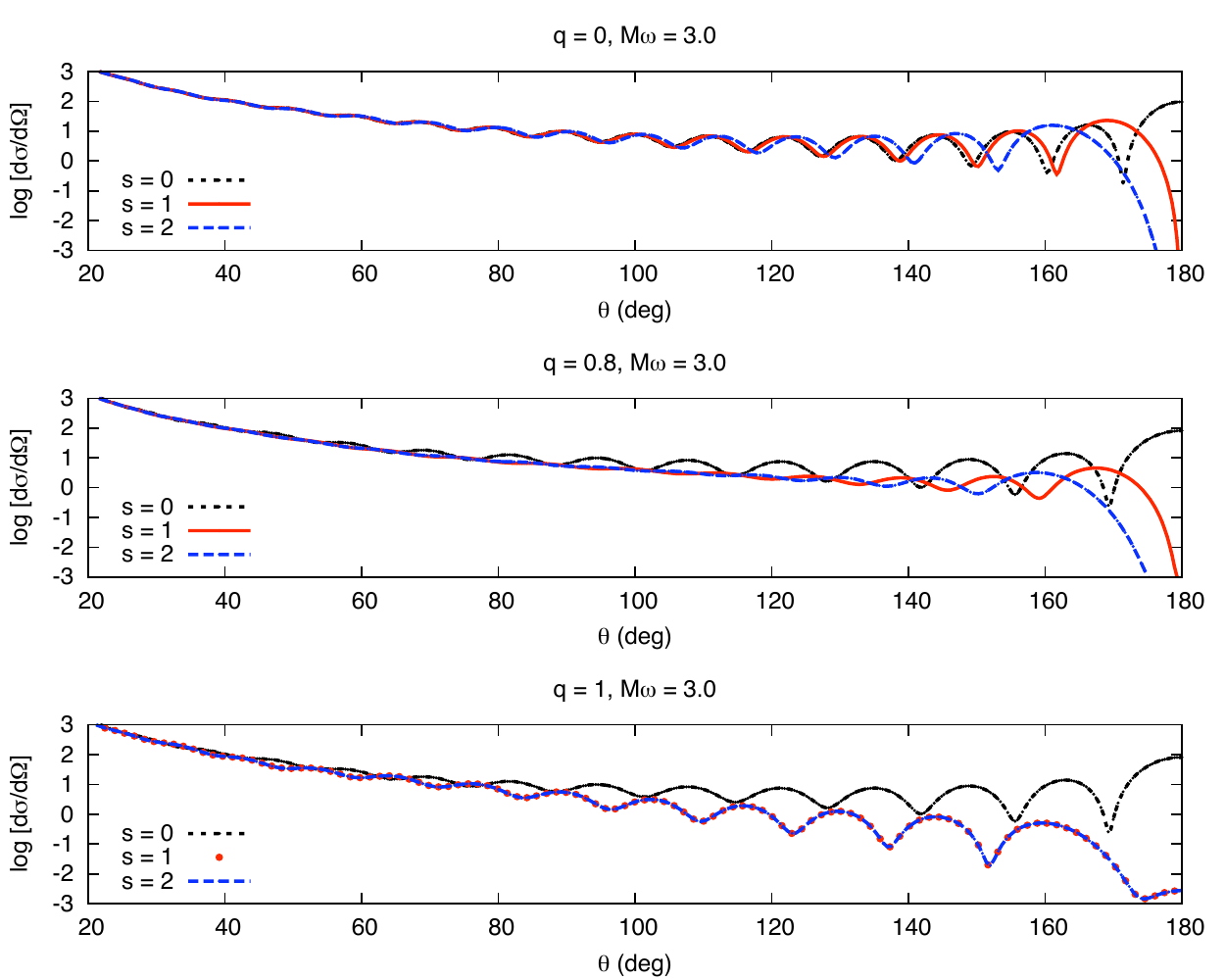}
 \caption{Scattering cross section for electromagnetic and
 gravitational radiation for Schwarzschild black holes (\emph{top}),
 Reissner-Nordstr\"om with $q \equiv |Q|/M = 0.8$ (\emph{middle}), and extreme
 Reissner-Nordstr\"om black holes (\emph{bottom}). Here,
 $M \omega = 3.0$.}
 \label{fig:rn_scs}
\end{figure}

Numerical investigations have revealed curious coincidences between electromagnetic and gravitational scattering 
properties in the extremal limit $Q \rightarrow M$. First, the resonant (i.e.~quasinormal mode) frequencies of 
gravitational waves with a multipole index $\ell$ coincide with those of electromagnetic waves with a multipole index $\ell-1$~\cite{Onozawa:1995vu}. Second, the absorption cross sections for gravitational and electromagnetic
waves are equal~\cite{Oliveira_2011:prd84_084048}. Third, as shown here for the first time,
the unpolarized (or circularly polarized) electromagnetic and gravitational scattering cross sections 
are equal (see Fig.~\ref{fig:rn_scs}, lower panel), and the 
electromagnetic-gravitational conversion cross sections are also equal. 

These properties all hint at an underlying symmetry. Remarkably, the symmetry may be revealed through the methods of $N=2$ supergravity: a (partial) realization of Einstein's dream of unifying electromagnetism 
and gravity~\cite{VanNieuwenhuizen:1981ae}. Supergravity, as the gauge theory of supersymmetry, pairs together
bosonic and fermionic fields. In the $N=2$ model, the electromagnetic ($s=1$) and gravitational ($s=2$) fields are joined by gravitino fields ($s=3/2$). The extremal RN black hole
is a Bogomol'nyi-Prasad-Sommerfeld (BPS) state in 
$N=2$ supergravity, and thus it possesses an unbroken supersymmetry~\cite{Gibbons:1982fy}. 
This supersymmetry has previously been shown to be responsible for the equality in quasinormal 
frequencies~\cite{Onozawa_1997:prd55_R4529}.  It also leads to some relations 
among partial-wave scattering amplitudes~\cite{Okamura:1997}.
These relations have been used to explain the equality in the gravitational and electromagnetic absorption
cross sections mentioned above.
Here, with a novel analysis that applies supersymmetry transformations directly to plane-wave
solutions,
we show that supersymmetry accounts for \emph{all} the additional coincidences noted above. 

\emph{Analysis:} Majumdar-Papapetrou geometries take the form~\cite{Hartle:1972ya}
\begin{equation}
ds^2 = u^{-2} dt^2 - u^2(dx^2+dy^2+dz^2), \label{maj-pap}
\end{equation}
where $ds^2 = g_{\mu \nu} dx^\mu dx^\nu$, $A_\mu = (u^{-1}, 0, 0, 0)$, and we adopt the signature ($+,-,-,-$). 
With this ansatz, the Einstein-Maxwell equations reduce to Laplace's equation $\nabla^2 u = 0$
and $\partial u/\partial t = 0$ on a flat 
background. The fundamental solution, $u = 1 + M/\rho$ with $\rho^2 = x^2 + y^2 + z^2$, is recognized as the 
extremal RN spacetime in isotropic coordinates. 

In $N=2$ supergravity, the gravitational and electromagnetic fields are paired with gravitino fields 
represented by two
Majorana spinors, $\psi_\mu^{(I)}$ ($I=1,2$). The Lagrangian $\mathcal{L}_{\text{SuGra}}(g_{\mu\nu}, A_\mu, \psi_\mu^{(I)})$ reduces to the standard Einstein-Maxwell Lagrangian $\mathcal{L}_{\text{EM}}(g_{\mu\nu}, A_\mu)$ when $\psi_\mu^{(I)} = 0$. 
Thus, RN black holes are solutions in $N=2$ supergravity.

The Lagrangian 
$\mathcal{L}_{\text{SuGra}}$ is invariant, up to a total divergence, under certain
gauge transformations that mix
$g_{\mu\nu}$, $A_\mu$ and $\psi_\mu^{(I)}$ ($N=2$ local supersymmetry).  
Since the extreme RN spacetime preserves some of the supersymmetry, one may generate
electrovacuum solutions that are small perturbations of this spacetime from 
$N=2$ supergravity solutions with small gravitino perturbations by applying supersymmetry transformations.

The local supersymmetry transformation for the suitably normalized metric perturbation $h_{\mu\nu}$  
and the electromagnetic field $A_\mu$ can be given as follows, up to a common constant
factor~\cite{VanNieuwenhuizen:1981ae}:
\begin{eqnarray} 
\delta A_\mu & = & \epsilon^{(1)T}\gamma^0 \psi_\mu^{(2)} - \epsilon^{(2)T}\gamma^0 \psi_\mu^{(1)},
\label{delta-A}\\
\delta h_{\mu\nu} & = & \frac{1}{\sqrt{2}}
( \epsilon^{(1)T}\gamma^0\gamma_{(\mu}\psi^{(1)}_{\nu)}
+ \epsilon^{(2)T}\gamma^0 \gamma_{(\nu}\psi_{\mu)}^{(2)}), \label{delta-h}
\end{eqnarray}
where $\epsilon^{(I)}$ ($I=1,2$) are Majorana spinors and ${}^T$ denotes the transpose. 
The $\gamma$-matrices satisfy  
$\{ \gamma^A,\gamma^B \} \equiv \gamma^A \gamma^B+ \gamma^B \gamma^A = - 2\eta^{A B}$, 
where $\eta^{AB}$ is the flat Lorentzian metric. We choose the $\gamma$-matrices 
to be real, noting that $\gamma^0$ is then anti-symmetric whereas $\gamma^A$ ($A=1,2,3$) are symmetric.  
We define
$\gamma_\mu = \gamma^A e_{A\mu}$, where the vectors $e_{A}^{\mu}$ $(A=0,1,2,3)$ form a local Lorentz
frame at each spacetime point. 
The supersymmetry
transformation of the gravitino fields can be given in terms of the field 
$\psi_\mu := \psi_\mu^{(1)} + i\psi_\mu^{(2)}$ and the gauge function 
$\epsilon := \epsilon^{(1)} + i\epsilon^{(2)}$ as follows~\cite{Gibbons:1982fy} 
with $F_{\mu\nu} = \partial_\mu A_\nu - \partial_\nu A_\mu$
on a background solution with $\psi^{(I)}_\mu=0$:
\begin{equation}
\delta \psi_\mu = D_\mu \epsilon 
+ \frac{i}{4} F_{\alpha\beta}\gamma^\alpha \gamma^\beta \gamma_\mu \epsilon, \label{delta-mu}
\end{equation}
where $D_\mu$ is the covariant derivative for spinors.

In Majumdar-Papapetrou spacetimes
there are Killing spinors $\epsilon$ satisfying $\delta\psi_\mu=0$ in Eq.~\eqref{delta-mu}.  
Any spinor $\epsilon= u^{-1/2}\epsilon_c$,
where $\epsilon_c$ is constant, is a Killing spinor if
$i\gamma^0\epsilon_c = -\epsilon_c$. This condition can be written in the real form as  
$\epsilon^{(2)} = -\gamma^0\epsilon^{(1)}$. 
Thus, the spinor $\epsilon^{(1)}$ can be chosen so that it is
any real constant spinor at spatial infinity.

The existence of a Killing spinor implies that $N=2$ supergravity theory linearized about the extreme 
RN spacetime is supersymmetric.  Therefore, we can map gravitino scattering solutions to
photon-graviton scattering solutions by supersymmetry.  More specifically, if $\psi_\mu^{(I)}$ give a
gravitino scattering solution, then from Eqs.~\eqref{delta-A} and \eqref{delta-h} and the relation
$\epsilon^{(2)}=-\gamma^0\epsilon^{(1)}$
we find that, if $\epsilon= \epsilon^{(1)} + i\epsilon^{(2)}$ is a Killing spinor, then
the following fields will give a photon-graviton scattering solution:
\begin{eqnarray}
A_\mu & = & \epsilon^{(1)T}(\psi_\mu^{(1)} + \gamma^0 \psi_\mu^{(2)}), \label{EMsol}\\
h_{\mu\nu} & = & \frac{1}{\sqrt{2}}(\epsilon^{(1)T}\gamma^0 \gamma_{(\mu} \psi_{\nu)}^{(1)}
- \epsilon^{(1)T}\gamma_{(\mu} \psi_{\nu)}^{(2)}).  \label{gravsol}
\end{eqnarray}
Since we are only interested in linearized field equations, we can treat $\psi_\mu^{(I)}$ and $\epsilon^{(I)}$
as ordinary functions of spacetime rather than fermionic operators or Grassmann numbers.

In order to discuss gravitino scattering solutions we first should consider gravitino plane-wave solutions in flat spacetime.
The non-interacting gravitino fields $\psi_\mu^{(I)}$ satisfy
$\gamma_{[\mu} \partial_\nu \psi_{\lambda]}^{(I)} = 0$ in flat space~\cite{Rarita:1941mf}.
These equations are invariant under the following gauge transformations:
$\psi_\mu^{(I)} \mapsto \psi_\mu^{(I)} + \partial_\mu \eta^{(I)}$, where $\eta^{(I)}$ are any
Majorana spinors. Let us consider a gravitino plane wave incident from $z=\infty$. 
Under the gauge condition $\gamma^\mu \psi_\mu^{(I)} = 0$ the general solutions to these field equations
are indeed plane waves and 
take the form $\psi_\mu^{(I)} = \psi_\mu^{(I,c)}e^{-ik(t+z)}$,
where $\psi_\mu^{(I,c)}$ are
constant spinor-vectors. The gauge invariance mentioned above
can be used to gauge away the components $\psi_0^{(I)}$ and $\psi_3^{(I)}$ (see, e.g., 
Refs.~\cite{Sterman:1977ds,VanNieuwenhuizen:1981ae}).  Then we find
$\gamma^0 \gamma^3 \psi_\mu^{(I)}  =  \psi_\mu^{(I)}$, $\mu=1,2$
and $\psi_2^{(I)}  =  \gamma^1 \gamma^2 \psi_1^{(I)}$.  
From now on we let $\psi_\mu^{(I)}$ be a scattering solution in the extreme RN background
which becomes a plane wave as described above in the limit $z\to \infty$.

For $z\to \infty$ the components with $\mu=1$ and $2$ of the electromagnetic plane wave in
Eq.~\eqref{EMsol} may be nonzero and are given by
\begin{eqnarray}
A_1 & \approx & \epsilon^{(1)T}(\psi_1^{(1)} + \gamma^0\psi_1^{(2)}), \label{A1-general}\\
A_2 & \approx & \epsilon^{(1)T}\gamma^1 \gamma^2(\psi_1^{(1)} + \gamma^0 \psi_1^{(2)}) ,
\label{A2-general}
\end{eqnarray}
where we have used the relation $\psi_2^{(I)} \approx \gamma^1\gamma^2\psi_1^{(I)}$.
We also find that, for $z\to\infty$, the $h_{\mu\nu}$ given by Eq.~\eqref{gravsol} is nonzero
only if $\mu$ or $\nu$ is either $1$ or $2$.  It can also be shown that the $h_{\mu 0}$ and $h_{\mu 3}$ are
of pure-gauge form, and hence can be disregarded in considering the flux of gravitational wave coming in from
$z=\infty$.
Hence, the only relevant components of the gravitational plane wave described by Eq.~\eqref{gravsol} are
found for large $z$ as
\begin{eqnarray}
h_{22} \approx - h_{11} & \approx & \frac{1}{\sqrt{2}}\epsilon^{(1)T}\gamma^0 \gamma^1 (\psi_1^{(1)}
- \gamma^0\psi_1^{(2)}),\label{h11h22-general}\\
h_{12} & \approx & 
- \frac{1}{\sqrt{2}}\epsilon^{(1)T}\gamma^0\gamma^2(\psi_1^{(1)} - \gamma^0 \psi_1^{(2)}).
\label{h12-general}
\end{eqnarray}
We note that these expressions and Eqs.~\eqref{A1-general} and \eqref{A2-general} become exact in the
limit $z\to\infty$, where $\psi_\mu^{(I)}$ gives a plane wave.

Now, we shall find gravitino plane-wave solutions that give 
linearly polarized incident electromagnetic plane waves through
Eq.~\eqref{EMsol} but no incident gravitational plane waves through Eq.~\eqref{gravsol}.  We first note that the 
$4$-dimensional space of constant real spinors consists of four $1$-dimensional eigenspaces of
the commuting real symmetric matrices $i\gamma_5\gamma^3$ and $\gamma^2$, each with eigenvalues $\pm 1$,
where $\gamma_5 := - i\gamma^0 \gamma^1\gamma^2\gamma^3$. We choose the scattering solution
$\psi_\mu^{(I)}$ such that $\psi_1^{(1)} + \gamma^0 \psi_1^{(2)}$
is an eigenspinor of $i\gamma_5\gamma^3$ and $\gamma^2$ with
eigenvalues $+1$ in the limit $z\to \infty$ where $\psi_\mu^{(I)}$ is an incoming plane wave.
Since $\psi^{(I)}_1 \approx \gamma^0\gamma^3\psi^{(I)}_1$ for $z\to\infty$,
we have
$\psi^{(1)}_1 - \gamma^0\psi^{(2)}_1 \approx \gamma^0 \gamma^3 (\psi_1^{(1)} + \gamma^0\psi_1^{(2)})$.
Thus,  $\psi^{(1)}_1 -\gamma^0\psi_1^{(2)}$ is 
an eigenspinor of $i\gamma_5\gamma^3$ with eigenvalue $-1$ for large $z$.
Define $\epsilon_{+\pm}$ to be unit real spinors satisfying
$i\gamma_5\gamma^3 \epsilon_{+\pm} = \epsilon_{+\pm}$ and
$\gamma^2 \epsilon_{+\pm} = \pm \epsilon_{+\pm}$.  Then, we find from 
Eqs.~\eqref{A1-general}-\eqref{A2-general} and Eqs.~\eqref{h11h22-general}-\eqref{h12-general} that
the incident electromagnetic wave is linearly polarized in the $1$st direction if 
$\epsilon^{(1)} = \epsilon_{++}$ and in the $2$nd direction if $\epsilon^{(1)} = \epsilon_{+-}$.  
In either case there will be no incident gravitational wave.

Let us now discuss the differential scattering and conversion cross sections for unpolarized 
incident electromagnetic waves~\footnote{Since the RN black holes are parity invariant, the cross sections for
circularly polarized waves are equal to those for unpolarized waves.}.  
Far away from the origin a scattered
solution can locally be approximated by a plane wave times $1/r$.   Now, choose a rotated coordinate system
characterized by the orthonormal vectors $\mathbf{e}_{\hat{i}}$, $i=1,2,3$, such that the unit vector in the
$3$rd direction in the old coordinate system, $\mathbf{e}_3$, is expressed as
$\mathbf{e}_3 = n_{\hat{1}}\mathbf{e}_{\hat{1}} + n_{\hat{2}}\mathbf{e}_{\hat{2}} 
+ n_{\hat{3}}\mathbf{e}_{\hat{3}}$.
After a suitable gauge transformation, the electromagnetic and gravitational plane waves in 
Eqs.~\eqref{EMsol} and \eqref{gravsol}, respectively,
propagating in the negative $\hat{3}$-direction are given by replacing the coordinates $1$, $2$ and $3$ by
$\hat{1}$, $\hat{2}$ and $\hat{3}$ in Eqs.~\eqref{A1-general}-\eqref{A2-general} and
Eqs.~\eqref{h11h22-general}-\eqref{h12-general}.
Hence the differential scattering 
cross section in the negative $\hat{3}$-direction is
\begin{eqnarray}
\left. \frac{d\sigma}{d\Omega}\right|_{A\to A}  
& = & C \sum_{\pm} \lim_{r\to\infty} r^2\left[ |A_{\hat{1}}|^2 + |A_{\hat{2}}|^2\right] \nonumber \\
& = & C\lim_{r\to\infty} r^2
(\psi_{\hat{1}}^{(1)\dagger}-\psi_{\hat{1}}^{(2)\dagger}\gamma^0)\nonumber \\
&& \times \left[ \Sigma_{+} + 
\gamma^{\hat{2}}\gamma^{\hat{1}}\Sigma_{+}
 \gamma^{\hat{1}}\gamma^{\hat{2}}\right](\psi_{\hat{1}}^{(1)} + \gamma^0\psi_{\hat{1}}^{(2)}), 
\nonumber\\
\label{Ps}
\end{eqnarray}
where $C$ is a constant and
$\Sigma_+   =  \epsilon_{++}\epsilon^T_{++} + \epsilon_{+-}\epsilon_{+-}^T 
\approx (1+i\gamma_5\gamma^3)/2$.
By substituting this formula into Eq.~\eqref{Ps} and using
$\gamma^3=n_{\hat{1}}\gamma^{\hat{1}} + n_{\hat{2}}\gamma^{\hat{2}} + n_{\hat{3}}\gamma^{\hat{3}}$,
we find
\begin{eqnarray}
\left. \frac{d\sigma}{d\Omega}\right|_{A\to A}
& = & C\lim_{r\to\infty} r^2(\psi_{\hat{1}}^{(1)\dagger} - \psi_{\hat{1}}^{(2)\dagger}\gamma^0)
\nonumber \\
&& \times (1 + in_{\hat{3}}\gamma_5 \gamma^{\hat{3}})(\psi_{\hat{1}}^{(1)} 
+ \gamma^0\psi_{\hat{1}}^{(2)}) \nonumber \\
& = &  \mathcal{A} + n_{\hat{3}}\mathcal{B},  \label{A-scattering}
\end{eqnarray}
where
\begin{eqnarray}
\mathcal{A}  & : = & C\lim_{r\to\infty} r^2\left[ \psi_{\hat{1}}^{(1)\dagger}\psi_{\hat{1}}^{(1)}
+ \psi_{\hat{1}}^{(2)\dagger}\psi_{\hat{1}}^{(2)}\right],\\
\mathcal{B} & : = & iC \lim_{r\to\infty} r^2\left[ \psi_{\hat{1}}^{(2)\dagger}\gamma_5\psi^{(1)}_{\hat{1}}
- \psi_{\hat{1}}^{(1)\dagger}\gamma_5 \psi^{(2)}_{\hat{1}}\right].
\end{eqnarray}
We have used the fact that $\psi_{\hat{1}}^{(I)}$ are eigenspinors of $\gamma^0 \gamma^{\hat{3}}$ with
eigenvalue $+1$.
The differential cross section in the same 
direction for the conversion from (unpolarized) 
electromagnetic wave to gravitational wave can be found similarly as
\begin{eqnarray}
\left. \frac{d\sigma}{d\Omega}\right|_{A\to h}
& = & C\sum_{\pm} \lim_{r\to\infty} r^2 \left[ |h_{\hat{1}\hat{1}}|^2 + |h_{\hat{2}\hat{2}}|^2
+ 2|h_{\hat{1}\hat{2}}|^2\right] \nonumber \\
& = & C\lim_{r\to\infty}r^2(\psi_{\hat{1}}^{(1)\dagger} + \psi_{\hat{1}}^{(2)\dagger}\gamma^0)
\nonumber \\
&& \times ( 1 + in_{\hat{3}}\gamma_5\gamma^{\hat{3}})
(\psi_{\hat{1}}^{(1)} - \gamma^0\psi_{\hat{1}}^{(2)}) \nonumber \\
& = & \mathcal{A} - n_{\hat{3}}\mathcal{B}. \label{Ah-conversion}
\end{eqnarray}

Next, we let $\epsilon^{(1)}=\epsilon_{-\pm}$, where $\epsilon_{-\pm}$ are real unit spinors satisfying
$i\gamma_5\gamma^3\epsilon_{-\pm} = - \epsilon_{-\pm}$ and
$\gamma^2 \epsilon_{-\pm} = \pm \epsilon_{-\pm}$.
Then, we find from 
Eqs.~\eqref{A1-general}-\eqref{A2-general} and Eqs.~\eqref{h11h22-general}-\eqref{h12-general}
that the solutions given by Eqs.~\eqref{EMsol} and \eqref{gravsol} are such that there is a linearly-polarized 
incident gravitational wave with the same flux as above coming in from $z=\infty$
with the `$+$' polarization (the `{\sf x}' polarization) if $\epsilon^{(1)} = \epsilon_{-+}$
($\epsilon^{(1)} = \epsilon_{--}$).  In either case there will be no incident electromagnetic wave.  
We can express the
unpolarized differential scattering cross section in the negative $\hat{3}$-direction as:
\begin{eqnarray}
\left. \frac{d\sigma}{d\Omega}\right|_{h\to h}
& = & 
C\sum_{\pm} 
\lim_{r\to\infty} r^2\left[ |h_{\hat{1}\hat{1}}|^2 + |h_{\hat{2}\hat{2}}|^2 + 2|h_{\hat{1}\hat{2}}|^2\right]
\nonumber \\
& = & \mathcal{A} + n_{\hat{3}}\mathcal{B},  \label{h-scattering}
\end{eqnarray}
where the summation is over $\epsilon_{-\pm}$.
We have used 
$\epsilon_{-+}\epsilon^{T}_{-+} + \epsilon_{--}\epsilon^T_{--} \approx (1 - i\gamma_5\gamma^3)/2$.
We similarly find the differential conversion cross section from 
gravitational to electromagnetic waves in the negative 
$\hat{3}$-direction as
\begin{eqnarray}
\left. \frac{d\sigma}{d\Omega}\right|_{h\to A}
& =& C\sum_{\pm}\lim_{r\to\infty} r^2\left[ |A_{\hat{1}}|^2 + |A_{\hat{2}}|^2\right] \nonumber \\
& = & \mathcal{A} - n_{\hat{3}}\mathcal{B}.  \label{hA-conversion}
\end{eqnarray}

By comparing Eqs.~\eqref{A-scattering} and \eqref{h-scattering} we see that the scattering cross sections for
the gravitational and electromagnetic waves are equal. We also find, by comparing Eqs.~\eqref{Ah-conversion}
and \eqref{hA-conversion}, that the conversion cross sections from electromagnetic to gravitational waves and
from gravitational to electromagnetic waves are equal.

\emph{Numerical results:} We now show that equality of scattering cross sections is supported by new numerical results, obtained via a rather different formalism: partial-wave solutions constructed from multipolar sums. 
For this purpose, 
we compute the scattering cross section of incident gravitational plane waves by RN black holes and  
compare it with the electromagnetic counterpart~\cite{Crispino_2014:prd90_064027}.
 
The multipoles of the electromagnetic and gravitational perturbations are
given by~\cite{Zerilli_1974:prd9_860}:
\begin{eqnarray}
 F_\ell^\mathcal{P} & = & \varphi_{\ell+}^\mathcal{P}\cos\psi -
 \varphi_{\ell-}^\mathcal{P}\sin\psi, \nonumber \\
 G_\ell^\mathcal{P} & = & \varphi_{\ell+}^\mathcal{P} \sin\psi +
 \varphi_{\ell-}^\mathcal{P} \cos\psi,
\end{eqnarray}
where the functions $\varphi_{\ell\pm}^\mathcal{P}$ obey the equations
\begin{equation}
 \frac{d^2}{dr_*^2}\varphi_{\ell\pm}^\mathcal{P} +
\left(\omega^2-V_{\ell\pm}^\mathcal{P}
 \right)\varphi_{\ell\pm}^\mathcal{P} = 0.
 \label{radial_eq}
\end{equation}
Here, $\mathcal{P}=\pm $ denotes parity (even [polar] or odd [axial]), 
$r_*$ is the tortoise coordinate, given by $d/dr_* \equiv f(r)\,d/dr$
with $f(r) = 1 -2M/r + Q^2/r^2$ in the standard coordinate system, and
\begin{equation}
 \sin(2\psi) = -2 \mathcal{P} Q\frac{\left[(\ell-1)(\ell+2) \right]^{1/2}}{\Omega},
 \qquad |\psi|<\frac{\pi}{4},
 \label{sin2psi}
\end{equation}
with $\Omega = \sqrt{ 9M^2 + 4Q^2(\ell-1)(\ell+2) }$.

The effective potentials, 
$V_{\ell\pm}^\mathcal{P}$, can be
found, for example, in Ref.~\cite{Moncrief_1974:prd10_1057}, with plots in Ref.~\cite{Chandra_1983}. 
As a general behavior, $V_{\ell\pm}^\mathcal{P}$ are zero at the
event horizon, $r_+$, and at infinity. Therefore, it is possible to
write:
\begin{equation}
 \varphi_{\ell\pm}^{\mathcal{P}} (r_*) \sim \left\{
\begin{array}{lr}
e^{-i\omega r_*} + A_{\pm,\omega \ell}^{\mathcal{P}}e^{i\omega r_*}, 
& (r_* \to \infty); \\
B_{\pm,\omega \ell}^{\mathcal{P}} e^{-i\omega r_*}, & (r_* \to - \infty).
\end{array}
\right.
\label{phi_asy}
\end{equation}

The gravitational scattering cross section for RN
black holes can be written as:
\begin{eqnarray}
\left. \frac{d\sigma}{d\Omega} \right|_{h \rightarrow h} &=& |\mathcal{F}|^2 + |\mathcal{G}|^2, 
\label{scs}
\end{eqnarray}
\begin{eqnarray}
 \mathcal{F} &=& \frac{\pi}{i\omega} \sum_{\mathcal{P} = \pm 1}
 \sum\limits_{\ell = 2}^{\infty} \, Z_\ell(1) Z_\ell (\cos\theta)
 \left[ e^{2i\delta_\ell^\mathcal{P}(\omega) } - 1 \right],
 \label{mathcal_F}
\\
 \mathcal{G} &=& \frac{\pi}{i\omega} \sum_{\mathcal{P} = \pm 1}
 \sum\limits_{\ell = 2}^{\infty} \mathcal{P} (-1)^{\ell}
 \, Z_\ell (1) Z_\ell (-\cos\theta) \nonumber \\
 && \quad \quad \quad \quad \quad \quad  \times \left[ e^{2i\delta_\ell^\mathcal{P}(\omega) } - 1 \right] ,
 \label{mathcal_G}
\end{eqnarray}
where $Z_\ell \equiv _{s=-2}Y_\ell^{m=2}(\cdot)$ are the spin-weighted spherical harmonics~\cite{Goldberg_1967:jmp_8_11},
and the phase shifts are given by
\begin{equation}
e^{2i\delta_\ell^\mathcal{P}(\omega)} = (-1)^{\ell+1}
(A_{-,\omega \ell}^\mathcal{P}\cos^2 \psi +
A_{+,\omega \ell}^\mathcal{P} \sin^2 \psi).
\label{A_ps_rel}
\end{equation}
The $\mathcal{F}$ and $\mathcal{G}$ are the
helicity-preserving and the helicity-reversing amplitudes,
respectively. We note that $\mathcal{G}$ is associated with the difference
of phase shifts from distinct parities and, therefore, is
zero whenever $\delta_\ell^+(\omega) = \delta_\ell^-(\omega)$. This is the case, e.g., for 
electromagnetic radiation around a Schwarzschild ($Q=0$) black holes, but not 
for a Reissner-Nordstr\"om black hole. Observational consequences of  this 
`helicity-nonconservation' were explored in Ref.~\cite{Crispino_2014:prd90_064027}. 

Figure \ref{fig:rn_scs} shows numerically-computed scattering cross sections for
waves of spin 0, 1, and 2 for black holes with $q = Q/M = 0, 0.8, 1$ and
$M \omega = 3.0$. The cross sections for Schwarzschild black holes are broadly
similar across the range $\theta \lesssim 100 ^\circ$, but differ at large angles. For $q = 0.8$,
results for $s = 1,2$ are very close for $\theta \lesssim 100^\circ$,
but differ significantly from the scalar case.
For extremal RN black holes ($q=1$), we see excellent numerical confirmation that the electromagnetic and gravitational scattering cross sections are equal (lower plot). Further numerical results will be presented in a companion paper.  

\bigskip

\acknowledgments

The authors would like to thank Gary Gibbons for useful correspondence and
Conselho Nacional de Desenvolvimento Cient\'ifico e Tecnol\'ogico (CNPq), 
Coordena\c{c}\~ao de Aperfei\c{c}oamento de Pessoal de N\'ivel Superior (CAPES), and 
Funda\c{c}\~ao Amaz\^onia de Amparo a Estudos e Pesquisas do Par\'a (FAPESPA) for partial financial support.
A.~H.\ also acknowledges partial support from the
Abdus Salam International Centre for Theoretical Physics
through the Visiting Scholar/Consultant Programme,
and the Universidade Federal do Par\'a (UFPA)
in Bel\'em for kind hospitality.


\end{document}